# Reasoning Under the Principle of Maximum Entropy for Modal Logics K45, KD45, and S5

Tivadar Papai
University of Rochester
papai@cs.rochester.edu

Henry Kautz
University of Rochester
kautz@cs.rochester.edu

Daniel Stefankovic
University of Rochester
stefanko@cs.rochester.edu

## ABSTRACT

We propose modal Markov logic as an extension of propositional Markov logic to reason under the principle of maximum entropy for modal logics *K45*, *KD45*, and *S5*. Analogous to propositional Markov logic, the knowledge base consists of weighted formulas, whose weights are learned from data. However, in contrast to Markov logic, in our framework we use the knowledge base to define a probability distribution over non-equivalent epistemic situations (pointed Kripke structures) rather than over atoms, and use this distribution to assign probabilities to modal formulas. As in all probabilistic representations, the central task in our framework is inference. Although the size of the state space grows doubly exponentially in the number of propositions in the domain, we provide an algorithm that scales only exponentially in the size of the knowledge base. Finally, we briefly discuss the case of languages with an infinite number of propositions.

## 1. INTRODUCTION

The central reasoning task for probabilistic logics is to infer the probability of a query formula given a knowledge base. One such logic is propositional Markov logic [4], where the knowledge base consists of weighted propositional formulas. While the weighted formulas define a probability distribution over possible worlds, and increasing the weight of a formula increases the probability mass assigned to worlds where the formula is true, the weights are not true probabilities. Weights can be learned from data or from assertions about the subjective probabilities of statements, or from both using data and explicit assertions of subjective probabilities [20]. In any case, the information obtained from the training data or from an expert can be interpreted as probabilities of the propositional formulas in the KB. Hence, propositional Markov logic defines the probability of formulas in two steps: first learn the weights of formulas in the KB given data and/or subjective probabilities of these propositional formulas, and second, use the learned parameters to infer the probability of query formulas. Out of all the possible distributions which satisfy the probabilistic constraints imposed by the training data or domain expert, the one defined by Markov logic networks is the maximum entropy distribution [17], which makes Markov logic an appealing choice. Markov logic is not the first framework that has been proposed for doing inference under the principle of maximum entropy. For example, a first-order logic language is used in [12, 2] to reason under maximum entropy and the maximum entropy distribution is found using conditional probability constraints in [7, 25].

One of the common approaches for combining probabilities and modal logic builds on a probability distribution defined over possible worlds [15, 22]. Although efficient inference algorithms for probabilistic modal logic have appeared in the past [23], they have been based on using a probabilistic Kripke structure that is explicitly given, not learned from data or assertions about the probabilities of formulas. In contrast, our approach generalizes maximum entropy reasoning for propositional logics to allow both the formulas in the knowledge base and the queries to be propositional modal logic formulas. *K45*, *KD45* and *S5* are the modal logics typically referred to as the logics of beliefs and knowledge. Zero-one laws have been established for such logics [14, 21], which can make probabilistic reasoning challenging if the state space is not chosen carefully; hence, we restrict our domain to be a finite set of epistemic situations (pointed Kripke structures, *i.e.*, Kripke structures with a distinguished real world state). The advantage of these modal logics is that to enumerate all the non-equivalent epistemic situations, it suffices to iterate over a finite set as long as our set of propositional formulas $\Omega$ is finite. Although the number of non-equivalent epistemic situations is finite in our problem formulation, their number grows $2^{O(2^{|\Omega|})}$. The main contributions of the paper are to show how one can reason under the principle of maximum entropy when simple propositional logic is replaced by either single agent modal logic *K45*, *KD45* or *S5*, and to provide an exact inference algorithm based on counting, which can drastically reduce the doubly exponential cost of a naively implemented inference algorithm to one singly exponential in the size of the knowledge base. We briefly discuss the case of languages with an infinite number of propositions.

## 2. BACKGROUND

### 2.1 Markov Logic

*Propositional Markov logic* [4] is a knowledge representation language that uses weighted propositional formulas to define probability distributions over truth assignments to propositions. A propositional *Markov logic network* consists of a knowledge base $KB = \{(w_i, F_i) | i = 1, \ldots, m\}$, where $w_i \in \mathbb{R}$ and $F_i$ is a propositional formula over a fixed set of propositions $\Omega = \{p_1, \ldots, p_{|\Omega|}\}$, and defines a probability distribution over truth assignments $X$ to $\Omega$ as follows:

$$\Pr(X = x) = \frac{1}{Z} \exp(\sum_i w_i f_i(x)) \,, \quad (1)$$

where $f_i(x) = 1$ if $F_i$ is true under $x$, otherwise $f_i(x) = 0$, and where $Z = \sum_{x \in \mathcal{X}} \exp(\sum_i w_i f_i(x))$ is the partition function, and $\mathcal{X}$ denotes the set of all possible truth assignments to $\Omega$, (i.e, $|\mathcal{X}| = 2^{|\Omega|}$). Note that, (1) defines an exponential family of distributions (see e.g. [26]). Exponential families have the property that for a given set of $f_i$ they describe the maximum entropy distribution that satisfies the set of consistent constraints $\mathbb{E}[f_i] = c_i$. Consistent here means that there exists a probability distribution that satisfies all the constraints simultaneously. We can interpret $c_i$ as the probability of the propositional formula being satisfied under a randomly chosen truth assignment $x$, hence (1) defines the maximum entropy distribution over the state space of truth assignments to the propositions with the constraints $\mathbb{E}[f_i] = c_i$. The probability of an arbitrary propositional formula $F$ over $\Omega$ is defined to be the probability of $F$ being true under a randomly chosen truth assignment $X$, i.e.:

$$\Pr(F) = \sum_{x \in \mathcal{X} : F \text{ is satisfied under } x} \Pr(X = x) = \mathbb{E}[f_i] \,. \quad (2)$$

## 2.2 Modal Logics K45, KD45 and S5

Modal logics *K45*, *KD45* and *S5* [3] extend propositional or first-order logic by adding a non-truth-functional sentential operators; we will again only discuss the propositional case here. We use the symbol $B$ to represent the modal operator in the language. Where $\alpha$ is a well formed sentence, then $B\alpha$ is a well formed sentence. For example, if we take $B$ to mean "the agent knows that", then the formula $Bp \vee B\neg p$ means that agent $i$ knows whether or not $p$ holds. Note that this is quite different from the tautology $Bp \vee \neg Bp$.

Different modal operators for concepts such as belief, knowledge, desire, obligation, *etc.*, can be specified by the *axiom schemas* that they satisfy. In this paper, we will consider only modal logics *K45*, *KD45*, and *S5*. The properties of each of these logics is the subset of the following axioms and rules [6]:

R1. From $\phi$ and $\phi \supset \psi$ infer $\psi$ (Modus ponens)

R2. From $\psi$ infer $B\psi$ (Knowledge Generalization)

A1. All tautologies of propositional calculus

A2. $(B\phi \wedge B(\phi \supset \psi)) \supset B\psi$ (Distribution Axiom)

A3. $B\phi \supset \phi$ (Knowledge Axiom)

A4. $B\phi \supset BB\phi$ (Positive Introspection Axiom)

A5. $\neg B\phi \supset B\neg B\phi$ (Negative Introspection Axiom)

A6. $\neg Bfalse$ (Consistency Axiom)

We get *K45* if we take R1, R2, A1, A2, A4, and A5. Besides the axioms of *K45*, *KD45* contains A6 and *S5* contains A3. *S5* is generally used to represent knowledge, and *KD45* beliefs. *K45* is similar to *KD45*; however, it allows believing in contradicting statements.

The common property of these logics is that every formula has an equivalent representation that has depth one, i.e., if $B\phi$ is a subformula then $\phi$ does not contain any other modal operators. In the rest of the paper we will always assume that we are only dealing with depth one modal formulas.

A Kripke structure over a set of propositions $\Omega$ is a tuple $\mathcal{M} = (S, \pi, \mathcal{K})$ where $S \neq \emptyset$ is the set of states, $\pi : S \to \mathcal{X}$, where $\mathcal{X}$ is the set of truth assignments over $\Omega$ and $\mathcal{K} \subseteq S \times S$. If $s \in S$ then for a propositional formula $F$, we have $\mathcal{M}, s \models F$ if $F$ is satisfied under $\pi(s)$. For a formula $BF$, we have $\mathcal{M}, s \models BF$ iff $\forall (s, r) \in \mathcal{K} : \mathcal{M}, r \models F$. Moreover, $\mathcal{M}, s \models F_1 \wedge F_2$ iff $\mathcal{M}, s \models F_1$ and $\mathcal{M}, s \models F_2$, and $\mathcal{M}, s \models \neg F$ iff $\mathcal{M}, s \not\models F$.

For each different modal logic, Kripke structures with different properties are associated. Reflexive, symmetric, and transitive relations (equivalence relations) are associated with modal operators that satisfy *S5*. Euclidean, serial, and transitive relations are associated with *KD45*. While Euclidean, and transitive relations are associated with *K45*. For a more detailed description of Kripke structures see, *e.g.*, [3, 6].

A Kripke structure with a distinguished state (generally denoting the real world) is called a pointed Kripke structure or (epistemic) situation, hence an epistemic situation $\sigma = (s, S, \pi, \mathcal{K})$ where $s \in S$. We call two epistemic situations $\sigma_1$ and $\sigma_2$ equivalent if for every formula $F$ we have $\sigma_1 \models F$ if and only if $\sigma_2 \models F$. Using this definition of equivalence, we can partition situations into equivalence classes.

We can enumerate all the non-equivalent epistemic situations for *K45*, *KD45* and *S5*, *i.e.*, we can select a member from each equivalence class by storing the worlds the agent considers possible and a distinguished real world state [6]. The Kripke structures have a fully connected sub-graph belonging to the possible worlds, and there is a special state $s$; in the case of *S5*, $s$ is included in the fully connected states, and in *KD45*, there is an outgoing arc from $s$ to every node representing a possible world. In both cases, the set of possible worlds is never empty. The difference between *KD45* and *K45* is that the set of possible worlds in *K45* can be empty. Let $\Sigma_{K45}$, $\Sigma_{KD45}$ and $\Sigma_{S5}$ denote the set of all possible situations we can construct using the previous descriptions for modal logics *K45*, *KD45* and *S5*, respectively. According to [6], if a formula is satisfiable it must be satisfiable in one of the situations in our $\Sigma$, and since not any two members of $\Sigma$ are equivalent it is enough to consider the members of $\Sigma$ to count every non-equivalent epistemic situations exactly once. For *K45*, *KD45*, and *S5*, if the set of propositions ($\Omega$) is fixed, then $|\Sigma_{K45}| = 2^{2^{|\Omega|}} 2^{|\Omega|}$, $|\Sigma_{KD45}| = (2^{2^{|\Omega|}} - 1) 2^{|\Omega|}$, $|\Sigma_{S5}| = 2^{2^{|\Omega|}-1} 2^{|\Omega|}$, respectively. In each case, the real world can be chosen from the possible $2^{|\Omega|}$ truth assignments. In *K45* the worlds the agent can consider to be possible can be any subset of all the possible worlds, *i.e.*, can have $2^{2^{|\Omega|}}$ values; in *KD45*, the subset cannot be empty; and in *S5*, since the real world must be considered possible, we can only pick a subset of the remaining truth assignments. We see that $|\Sigma| \leq 2^{|\Omega|} 2^{2^{|\Omega|}}$ in all the three cases. We will denote the above mentioned sets by $\Sigma_\mathcal{L}(\Omega)$, where $\mathcal{L} \in \{K45, KD45, S5\}$, when we want to emphasize their dependence on $\Omega$.

## 3. DEFINING THE MAXIMUM ENTROPY DISTRIBUTION

Since the maximum entropy distribution can be sensitive to the choice of the state space (see, *e.g.*, [13, 16]), we have to be careful when we choose our state space in order to avoid non-intuitive results. *E.g.*, if $\Omega = \{p\}$, and we want to reason about the knowledge of someone using modal logic *S5*, a straightforward extension might seem to be to add a "modal atom" $Bp$ to $\Omega$ and define a probability distribution over the modally consistent truth assignments to this set $\{p, Bp\}$, ruling out *e.g.*, the case when $p$ is assigned *false* and $Bp$ is assigned *true*. However, it is easy to see that with an empty $KB$, $\Pr(p) = \frac{1}{3}$, which seems counter-intuitive, since we have no reason to believe that $p$ is more likely to be true than to be false (analogous examples in a different domain are given in [13]). Moreover, if $\Omega$ contains more propositions, selection of modal atoms becomes more complicated, *e.g.*, if $\Omega = \{p, q\}$ should we choose only $Bp$, $Bq$, $B\neg p$ and $B\neg q$ as modal atoms, or should we also include $B(p \vee q)$? Without including the latter, its probability can only be bounded but not determined, because a truth assignment to the rest of the modal atoms would not be sufficient to decide its truth value.

Based on the above mentioned problems our goals should be as follows:

(i) Assign probabilities to arbitrary modal or non-modal formulas over a fixed set of propositions $\Omega$ based on a set of weighted formulas $KB = \{(w_i, F_i)\}$ in a well-defined way.

(ii) If $KB$ contains only weighted non-modal formulas, we should obtain the distribution that propositional Markov logic would define.

(iii) If $KB$ does not contain infinite weights and $\psi$ subsumes $\phi$, and $\phi$ and $\psi$ are non-equivalent, then $\Pr(\psi) < \Pr(\phi)$.

These criteria can be achieved by assigning probabilities to epistemic situations rather than to modal atoms. Given a non-empty set of epistemic situations $\Sigma$ over a fixed $\Omega$ propositions, we define the probability of $\sigma \in \Sigma$ as:

$$\Pr(\sigma) = \frac{1}{Z} \exp(\sum_{i:\sigma \models f_i} w_i) , \qquad (3)$$

where the partition function is defined as:

$$Z = \sum_{\sigma \in \Sigma} \exp(\sum_{i:\sigma \models f_i} w_i) . \qquad (4)$$

The probability of a formula $\phi$ (modal or non-modal) is defined as:

$$\Pr(\phi) = \sum_{\sigma \in \Sigma: \sigma \models \phi} \Pr(\sigma) . \qquad (5)$$

Property (i) clearly holds, no matter how we choose $\Sigma$. To satisfy Property (ii) it must be true that $c(x) = |\{(\mathcal{M}, s) \in \Sigma | \pi(s) = x\}|$ has the same value for every truth assignment $x$ over $\Omega$. If $\Sigma$ contains every non-equivalent situations then Property (iii) is clearly satisfied as well. Hence, if we choose the state space to be $\Sigma_{K45}$, $\Sigma_{KD45}$ or $\Sigma_{S5}$, all the desired three conditions are satisfied.

Note that we could define the same distribution using modal atoms as we do by defining distribution over $\Sigma_{K45}$, $\Sigma_{KD45}$ or $\Sigma_{S5}$. *E.g.*, we define the same distribution if we choose the modal atoms to be all the propositional atoms, and all the depth one formulas in the form $Bc$, where $c$ is a conjunction which contains every proposition either as positive or a negative literal. However, for our goals we found the approach to define the distribution over epistemic situations more general, because in this way Property (i) always holds, we do not have to account for modally inconsistent states, moreover, it is easier to decide whether Properties (ii) and (iii) hold.

## 4. INFERENCE

The computationally expensive part of determining (3) and (5) can both be reduced to the computation of a partition function (4). *E.g.*, to infer the probability of a formula $F$ not present in the knowledge base $KB$, we first have to create a new knowledge base $KB' = KB \cup \{(\infty, F)\}$. If $Z$ and $Z'$ denote the partition functions corresponding to the knowledge bases $KB$ and $KB'$, respectively, then it follows from (3), (4) and (5) that $\Pr(F) = \frac{Z'}{Z}$.

Computing the partition function is challenging because the size of the state space for *K45*, *KD45* and *S5* are all doubly exponential in $|\Omega|$ as mentioned in Sec. 2.2. On the other hand, there is much symmetry in the domain, *i.e.*, many situations have the same probability; hence $\Sigma$ can be divided into equivalence classes. Similar to lifted inference techniques for quantified probabilistic logics (a highly active research area today, *e.g.* [24, 18, 11]), we show how one can compute the partition function without explicitly iterating through every state in the domain. Although our exact inference algorithm is exponential in a quantity describing the complexity of the knowledge base, it is vastly faster than iterating through the $2^{O(2^{|\Omega|})}$ epistemic situations in $\Sigma_{K45}$, $\Sigma_{KD45}$, or $\Sigma_{S5}$. We are going to assume that we have access to a propositional model counter, *i.e.*, for any propositional formula we can tell the number of its satisfying truth assignments. (Exhaustive solvers run in exponential time, which is sufficient for our claimed bounds, but heuristic/approximate solvers such as, *e.g.*, SampleSearch [10], may be more useful in practice.) We now show how to reduce the computation of a partition function to counting epistemic situations that satisfy a given set of modal logic formulas. We first introduce truth assignments to formulas in the knowledge base. If $KB = \{(w_1, F_1), \ldots, (w_n, F_n)\}$, let $\mathcal{T}$ be the set of length $n$ Boolean vectors. For $t \in \mathcal{T}$ let $\Phi(t)$ be a conjunction where the $i$-th term is $F_i$ if $t_i = \text{true}$, and it is $\neg F_i$ if $t_i = \text{false}$. Members of $\mathcal{T}$ will partition the space of epistemic situations $\Sigma$ into disjoint sets. $\sigma_1, \sigma_2 \in \Sigma$ will be in the same partition if for every $t \in \mathcal{T}$ we have $\sigma_1 \models \Phi(t)$ iff $\sigma_2 \models \Phi(t)$. If $\sigma_1$ and $\sigma_2$ are in the same partition then $\Pr(\sigma_1) = \Pr(\sigma_2)$. To simplify notation, let $w(t) = \sum_{i:t_i=\text{true}} w_i$, *i.e.*, $w(t)$ is the sum of the weights of the formulas to which $t$ assigns true. Hence, we can rewrite (4) as:

$$Z = \sum_{t \in \mathcal{T}} N(\Phi(t)) \exp(w(t)) , \qquad (6)$$

where $N(\phi)$ denotes the number of epistemic situations where $\phi$ holds, *i.e.*, $N(\phi) = |\{\sigma \in \Sigma | \sigma \models \phi\}|$.

The probability of any query formula $F$ can be written as:

$$\Pr(F) = \frac{1}{Z} \sum_{t \in \mathcal{T}} N(\Phi(t) \wedge F) \exp(w(t)) . \qquad (7)$$

Next, we show how to compute $N(F)$ for different formulas. Table 1 contains the simple counts for different basic formulas in *K45*, *KD45*, or *S5*. We use the notation $c(\phi)$ for the number of satisfying truth assignments of propositional formula $\phi$. Using the rules in (8) we can compute $N(F)$ for any formula which is in CNF normal form, where each term is either a propositional formula, or in the form $B\phi$ or $\neg B\phi$, where $\phi$ is a propositional formula. The most general form of a conjunction is $C = \phi_0 \wedge B\psi \wedge (\wedge_{i=1}^k \neg B\phi_i)$ (we call such conjunctions simple), since $B\phi_1 \wedge B\phi_2 = B(\phi_1 \wedge \phi_2)$. The counting of the satisfying assignments of $C$ is done by the inclusion-exclusion principle and by counting the members of the complement of sets.

EXAMPLE 1. *If $p$ and $q$ are propositions and $F = (p \supset q) \wedge B(p \vee q) \wedge \neg Bp \wedge \neg Bq$:*

$$\begin{aligned}
N(F) &= N((p \supset q) \wedge B(p \vee q) \wedge \neg Bp \wedge \neg Bq) \\
&= N((p \supset q) \wedge B(p \vee q)) - \\
&\quad [N((p \supset q) \wedge B(p \vee q) \wedge Bp) + \\
&\quad\phantom{[} N((p \supset q) \wedge B(p \vee q) \wedge Bq) - \\
&\quad\phantom{[} N((p \supset q) \wedge B(p \vee q) \wedge Bp \wedge Bq)] \\
&= N((p \supset q) \wedge B(p \vee q)) - N((p \supset q) \wedge Bp) - \\
&\quad N((p \supset q) \wedge Bq) + N((p \supset q) \wedge B(p \wedge q))
\end{aligned}$$

Hence, a CNF formula with this general type of conjunctions can represent any modal formula in *K45*, *KD45*, and *S5* since every formula in *K45*, *KD45*, and *S5* have an equivalent depth one representation (which can possibly increase the size of the formula drastically). (To see why a depth one representation for every formula $F$ exists, consider the set of modal atoms presented at the end of Sec. 3. Since every epistemic situation can be characterized by a conjunction of these modal atoms where each literals is either a positive or negative modal atom, we can form a depth one formula by taking the disjunction of the situations where $F$ holds.) The final piece of computation of $N(F)$ for a CNF formula $F$ again uses the inclusion-exclusion principle, replacing the computation for disjunctions with (exponentially) many conjunctions.

EXAMPLE 2. *In modal logic S5, $p$ and $q$ being propositions, the use of inclusion-exclusion principle to reduce the computation of $N(F)$ for the CNF formula $F = (p \supset q) \vee Bp) \wedge (p \vee B(p \vee q))$ proceeds as follows:*

$$\begin{aligned}
N(F) &= N(((p \supset q) \vee Bp) \wedge (p \vee B(p \vee q))) \\
&= N((p \supset q) \wedge (p \vee B(p \vee q))) + \\
&\quad N(Bp \wedge (p \vee B(p \vee q))) - \\
&\quad N((p \supset q) \wedge Bp \wedge (p \vee B(p \vee q))) \\
&= N((p \supset q) \wedge p) + N((p \supset q) \wedge B(p \vee q)) - \\
&\quad N((p \supset q) \wedge p \wedge B(p \vee q)) + N((Bp \wedge p) + \\
&\quad N(Bp \wedge B(p \vee q)) - N(Bp \wedge p \wedge B(p \vee q)) - \\
&\quad (N((p \supset q) \wedge Bp \wedge p) + \\
&\quad N((p \supset q) \wedge Bp \wedge B(p \vee q)) - \\
&\quad N((p \supset q) \wedge Bp \wedge p \wedge B(p \vee q)))
\end{aligned}$$

*We see that each term can be easily computed using the rules in Table 1 and the expressions in (8).*

$$N(\vee_{i=1}^k B\phi_i) = \sum_{i=1}^k (-1)^{i+1} \sum_{1 \leq j_1 < j_2 < \ldots < j_i \leq k} N(B(\phi_{j_1} \wedge \ldots \wedge \phi_{j_i}))$$

$$N(\wedge_{i=1}^k \neg B\phi_i) = N(true) - \sum_{i=1}^k (-1)^{i+1} \sum_{1 \leq j_1 < j_2 < \ldots < j_i \leq k} N(B(\phi_{j_1} \wedge \ldots \wedge \phi_{j_i}))$$

$$N(\phi_0 \wedge B\psi \wedge (\wedge_{i=1}^k \neg B\phi_i)) = N(\phi_0 \wedge B\psi) - \sum_{i=1}^k (-1)^{i+1} \sum_{1 \leq j_1 < j_2 < \ldots < j_i \leq k} N(\phi_0 \wedge B\psi \wedge B(\phi_{j_1} \wedge \ldots \wedge \phi_{j_i}))$$

$$N((\vee_{i=1}^k F_i) \wedge F) = \sum_{i=1}^k (-1)^{i+1} \sum_{1 \leq j_1 < j_2 < \ldots < j_i \leq k} N(F \wedge F_{j_1} \wedge \ldots \wedge F_{j_i})$$

(8)

Using the established rules of counting we could give a time complexity result of our inference algorithm for CNF formulas, but instead we give results for formulas in a more general form. The most general language $\mathcal{L}(\Omega)$ we use is defined as follows:

- Every propositional formula is a member of $\mathcal{L}(\Omega)$,
- If $\phi$ is a propositional formula, then $B\phi \in \mathcal{L}(\Omega)$,
- If $\phi \in \mathcal{L}(\Omega)$, then $\neg\phi \in \mathcal{L}(\Omega)$,
- If $\phi_1, \phi_2 \in \mathcal{L}(\Omega)$, then $\phi_1 \wedge \phi_2 \in \mathcal{L}(\Omega)$.

Hence, we only allow depth one modal formulas, however, since in modal logics *K45*, *KD45*, and *S5*, every formula has an equivalent depth one representation, we can allow this restriction without the loss of generality.

THEOREM 1. *Counting the non-equivalent epistemic situations that satisfy a depth one formula $F \in \mathcal{L}(\Omega)$ in K45, KD45, or S5 can be accomplished in time $2^{O(|F|+|\Omega|)}$.*

We use the following definition and lemma to prove Theorem 1.

DEFINITION 1. *For a formula $F$ let $\mathbb{I}_F : \Sigma \to \{0,1\}$ denote the characteristic function of $F$ in the space of all non-equivalent epistemic situations, i.e., $\mathbb{I}_F(\sigma) = 1$ iff $\sigma \models F$.*

The next result shows that the characteristic function of a depth one formula can be expressed as a combination of characteristic functions of simple conjunctions.

LEMMA 1. *For a depth one modal logic formula $F \in \mathcal{L}(\Omega)$ in $K45, KD45$ or $S5$, it is always possible to represent $\mathbb{I}_F = \sum_{i=1}^K w_i \mathbb{I}_{C_i}$ where every $w_i \in \{-1, +1\}$, $|C_i| \leq |F|$, every $C_i$ is a simple conjunction (i.e., in the form $\phi_0^i \wedge B\psi^i \wedge (\wedge_{k=1}^{l_i} \neg B\phi_k^i)$), and $K \leq 2^{|F|}$.*

PROOF. We prove the lemma by induction on the structure of $F$. For the base cases where $F$ is either a propositional formula, or in the form $B\phi$ where $\phi$ is a propositional formula, the claim clearly holds. Suppose that $F = \neg F_1$.

| F | $N(F)$ | | |
|---|---|---|---|
| | K45 | KD45 | S5 |
| true | $|\Sigma_{K45}| = 2^{|\Omega|}2^{2^{|\Omega|}}$ | $|\Sigma_{KD45}| = 2^{|\Omega|}(2^{2^{|\Omega|}} - 1)$ | $|\Sigma_{S5}| = 2^{|\Omega|}2^{2^{|\Omega|}-1}$ |
| propositional formula $\phi$ | $c(\phi)2^{2^{|\Omega|}}$ | $c(\phi)(2^{2^{|\Omega|}} - 1)$ | $c(\phi)2^{2^{|\Omega|}-1}$ |
| $B\phi$ | $2^{|\Omega|}2^{c(\phi)}$ | $2^{|\Omega|}(2^{c(\phi)} - 1)$ | $c(\phi)2^{c(\phi)-1}$ |
| $\phi_0 \wedge B\phi$ | $c(\phi_0)2^{c(\phi)}$ | $c(\phi_0)(2^{c(\phi)} - 1)$ | $c(\phi_0 \wedge \phi)2^{c(\phi)-1}$ |
| $\neg B\phi$ | $|\Sigma_{K45}| - N(B\phi)$ | $|\Sigma_{KD45}| - N(B\phi)$ | $|\Sigma_{S5}| - N(B\phi)$ |

Table 1: Basic counting rules for *K45*, *KD45*, and *S5*

The claim of the lemma holds for $F_1$ (by induction hypothesis), and hence:

$$\mathbb{I}_{F_1} = \sum_{i=1}^{K} w_i \mathbb{I}_{C_i} . \tag{9}$$

Then we have

$$\mathbb{I}_F = \mathbb{I}_{true} - \mathbb{I}_{F_1} = \mathbb{I}_{true} - \sum_{i=1}^{K} w_i \mathbb{I}_{C_i} . \tag{10}$$

Now suppose $F = F_1 \wedge F_2$. Then:

$$\mathbb{I}_F = \mathbb{I}_{F_1} \mathbb{I}_{F_2} = \sum_{i=1}^{K_1}\sum_{j=1}^{K_2} w_i^1 w_j^2 \mathbb{I}_{C_i^1} \mathbb{I}_{C_j^2} = \sum_{i=1}^{K_1}\sum_{j=1}^{K_2} w_i^1 w_j^2 \mathbb{I}_{C_i^1 \wedge C_j^2} . \tag{11}$$

Notice that $|C_1 \wedge C_2| \leq |F_1| + |F_2| + 1 \leq |F|$ and $K_1 K_2 \leq 2^{|F_1|+|F_2|} \leq 2^{|F|}$. $\square$

COROLLARY 1. *Since $N(F) = \sum_{\sigma \in \Sigma} \mathbb{I}_F(\sigma)$, we have*

$$N(F) = \sum_{i=1}^{K} \sum_{\sigma \in \Sigma} \mathbb{I}_{C_i}(\sigma) = \sum_{i=1}^{K} w_i N(C_i) , \tag{12}$$

*i.e., the problem of counting epistemic situations in which $F$ is satisfied has been reduced to counting epistemic situations in which the basic conjunctions $C_i$ are satisfied.*

Now we are ready to prove Theorem 1

PROOF PROOF OF THEOREM 1. According to Corollary 1, $N(F) = \sum_{i=1}^{K} w_i N(C_i)$. We first note that using the inclusion-exclusion principle (see (8)) for every $i$ we can compute $N(C_i)$ in time $2^{O(|C_i|+|\Omega|)}$. (Note that the bound on the running time is large enough to allow counting the satisfying assignments of the necessary propositional formulas.) Since $K \leq 2^{|F|}$ and for every $i$ we have $|C_i| \leq |F|$, computing $N(F)$ can be accomplished in $2^{O(|F|+|\Omega|)}$. $\square$

COROLLARY 2. *Computing the partition function in (6) for a knowledge base consisting of formulas with depth at most one can be accomplished in time $2^{O(|F|+|\Omega|)}$.*

## 5. INFINITE DOMAINS

Although the main focus of the paper is finite domains, we briefly discuss here the case of infinite domains. The source of finiteness in our formulation is that there are only a finite set of non-equivalent epistemic situations over a given set of propositions ($\Omega$). We now consider the questions regarding the effect of increasing the size of $\Omega$ where the state space is, as before, the set of non-equivalent epistemic situations:

1. Do zero-one laws hold for infinite domains?
2. Given a knowledge base of equality constraints on the probabilities of formulas, are there formulas the probability of which have to be either 0 or 1?

The existence of zero-one laws is well-known for first-order logic [9, 5] and for modal logic [14]. In the modal logic setting, the zero-one law states that given an arbitrary formula, the probability of it being valid in a randomly chosen Kripke structure with $N$ number of states converges to 1 or to 0 as $N \to \infty$. In [14], the state space can contain multiple Kripke structures with $N$ states that are equivalent; hence, the size of the state space is not bounded, despite $\Omega$ being finite. Moreover, the focus of their paper is on the probability of a formula being valid in a randomly chosen Kripke structure, while we are interested in the probability of a formula being satisfied in a randomly chosen epistemic situation. To show the contrast, consider the case of an empty knowledge base which defines a uniform distribution over the situations. The probability of a proposition $p$ being true will always be 0.5 regardless of $|\Omega|$, hence its probability is not going to converge to 0 or to 1. However, *e.g.* $\Pr(B\phi) \to 0$ if $\phi$ is not a tautology, otherwise $\Pr(B\phi) \to 1$ (we can verify this by taking the limit of $\frac{N(true)}{N(B\phi)}$ using Table 1 and that adding $k$ more propositions to $\Omega$ changes the value of $c(\phi)$ to $c(\phi)2^k$). More generally:

THEOREM 2. *If $C$ is a consistent simple conjunction, i.e., $C = \phi_0 \wedge B\psi \wedge (\wedge_{i=1}^{k} \neg B\phi_i)$ where $\psi$ and every $\phi_i$ is a propositional formula and $\beta$ is a propositional formula s.t. $C \wedge B\beta$ is consistent as well, then $\lim_{|\Omega| \to \infty} \frac{N(B\beta \wedge C)}{N(C)} = 0$ if $\psi \not\models \beta$, otherwise $\lim_{|\Omega| \to \infty} \frac{N(B\beta \wedge C)}{N(C)} = 1$.*

The proof of Theorem 2 makes use of the following lemmas (which we prove only for *K45*).

LEMMA 2. *If $\phi_0$, $\psi$ and $\beta$ are propositional formulas and $\phi_0 \wedge \psi$ is satisfiable then $\lim_{|\Omega| \to \infty} \frac{N(\phi_0 \wedge B\psi \wedge B\beta)}{N(\phi_0 \wedge B\psi)} = 0$ if $\psi \not\models \beta$, otherwise $\lim_{|\Omega| \to \infty} \frac{N(\phi_0 \wedge B\psi \wedge B\beta)}{N(\phi_0 \wedge B\psi)} = 1$.*

PROOF. We only prove the lemma for *K45*. Similar proof works for *KD45* and *S5*. Assume $\phi_0$, $\psi$ and $\beta$ only build on propositions from a set $\Omega'$ and let $k = |\Omega| - |\Omega'|$. For a propositional formula $F$ that builds only on propositions from $\Omega'$ let $c'(F)$ denote the number of its satisfying truth assignments over $\Omega'$. We have

$$\frac{N(\phi_0 \wedge B\psi \wedge B\beta)}{N(\phi_0 \wedge B\psi)} = \frac{2^k c'(\phi_0 \wedge \psi \wedge \beta) 2^{2^k c'(\psi \wedge \beta)}}{2^k c'(\phi_0 \wedge \psi) 2^{2^k c'(\psi)}} \tag{13}$$

$$= \frac{c'(\phi_0 \wedge \psi \wedge \beta)}{c'(\phi_0 \wedge \psi)} 2^{2^k (c'(\psi \wedge \beta) - c'(\psi))}$$

(note that $N$ counts epistemic situations over $\Omega$ whereas $c'$ counts satisfying assignments over $\Omega'$). Since if $\psi \not\models \beta$ then $c'(\psi \wedge \beta) - c'(\phi) < 0$, hence the ratio converges to 0 as $k \to \infty$. It is easy to verify that if $\psi \models \beta$ this ratio is 1 for every $k$. □

The next result means that as the number of extra propositions increases we can remove terms in the form $\neg B\phi$ from simple conjunctions.

LEMMA 3. *If $C$ is a consistent simple conjunction, i.e., $C = \phi_0 \wedge B\psi \wedge (\wedge_{i=1}^{k} \neg B\phi_i)$ where $\psi$ and every $\phi_i$ is a propositional formula then $\lim_{|\Omega| \to \infty} \frac{N(\phi_0 \wedge B\psi)}{N(C)} = 1$.*

PROOF. If we expand $C$ according to the inclusion-exclusion principle (equation (8)) we can conclude that

$$\lim_{|\Omega| \to \infty} \frac{N(\phi_0 \wedge B\psi)}{N(C)} = 1$$

since for all the other terms

$$\lim_{|\Omega| \to \infty} \frac{N(\phi_0 \wedge B\psi \wedge B(\phi_{j_1} \wedge \ldots \wedge \phi_{j_i}))}{N(\phi_0 \wedge B\psi)} = 0$$

according to Lemma 2. □

We can prove now Theorem 2.

PROOF OF THEOREM 2. Theorem 2 immediately follows from the following telescopic product

$$\frac{N(B\beta \wedge C)}{N(\phi_0 \wedge B\psi \wedge B\beta)} \frac{N(\phi_0 \wedge B\psi \wedge B\beta)}{N(\phi_0 \wedge B\psi)} \frac{N(\phi_0 \wedge B\psi)}{N(C)} \quad (14)$$

where the first and third terms converge to 1 (using Lemma 3). The claim now follows from Lemma 2 applied to the second term. □

Using Theorem 2, we can simplify the computation of $N(\phi)$ in the limit $|\Omega| \to \infty$ for any formula $\phi$ by dropping terms in the form $\neg B\phi$ whenever we encounter a conjunction in the most general form; in addition, we can expect to neglect the majority of the terms when using the inclusion-exclusion rule. Unfortunately, one consequence of Theorem 2 is that the weight of certain formulas in the knowledge base will go to infinity as $|\Omega| \to \infty$. Consider, e.g., the simple formula $Bp$ in the knowledge base. If its weight is $w$, then $\Pr(Bp) = \frac{N(Bp) \exp(w)}{N(Bp) \exp(w) + N(\neg Bp)}$ which converges to 0 for any finite $w$, hence $0 < \Pr(Bp) < 1$ cannot be captured with any finite $w$ as we increase $|\Omega|$.

One way we can avoid this phenomenon is to define $w$ as a function of $|\Omega|$ as [16] suggests for first-order Markov logic. Another approach would be to choose $\Sigma$ not to include every non-equivalent episetemic situations. If this were done, any learned model would not be able to satisfy Property (iii), but could achieve non-zero and non-one probabilities for every modal formula with finite $w$ values.

## 6. RELATED WORK

Reasoning with a knowledge base of statistical information have been approached in many different ways. The ones making use of the principle of maximum entropy [17] seem to be more natural, because when multiple distributions are consistent with our knowledge base, then there is no reason to prefer one over the other. In [12], first-order logic is the representational language and a connection between maximum entropy reasoning is presented when unary predicates are used. Markov logic [4] is one of the most popular choices in the statistical relational learning community for reasoning under the maximum entropy with a first-order logic knowledge base. Propositional Markov logic is generalized in [7] by using different features that are capable of capturing conditional probabilities. Although we do not mention representation of conditional probabilities, our framework could be generalized in this direction. Maximum entropy models are sensitive to the choice of domain, but whether this is a property of other kinds of models is discussed in [13]. [16] proposes to make the weights dependent on the size of the domain to counter act against the change of marginals when the domain size changes. Hence, it is not surprising that we eventually encountered the issue of changing marginals in both of our chosen state spaces. Zero-one laws for first-order logic are long known [9, 5]; for modal logics, they were established in [14]; and for conditional probabilities in [21]. In our setting when we have a finite number of propositions our state space is always finite, hence we only experience the convergence of the probability of certain formulas to 0 and to 1 when we started increasing the number of propositions. Probabilistic modal logic has been proposed in [22] and an efficient inference algorithm in [23]. Although their proposed framework is capable of answering queries using given a probabilistic Kripke structure, but not suitable for learning the probabilistic model given a probabilistic knowledge base. In contrast, our approach defines an exponential family or probability distributions, hence the learning of the parameters of the distribution is a convex optimization problem (see e.g. [19]).

## 7. CONCLUSIONS AND FUTURE WORK

In this paper we showed a way to extend propositional Markov logic with modal operators using epistemic situations (pointed Kripke structures) as basic building blocks of the domain. The modal logics we focused on were *K45*, *KD45*, and *S5* for a single agent. The common theme in the modal logics we considered is that the number of non-equivalent epistemic situations is finite, but grows doubly exponentially with the number of propositions in the domain. However, we provided an exact inference algorithm where complexity is only singly exponential in the size of the knowledge base.

Although we only provided an exact inference algorithm, the three main parts of computations we need to perform for exact inference can all be approximated. Bonferroni inequalities can provide an approximation when we use the inclusion-exclusion principle in our computations; in addition, our discussion of infinite domains suggest that the bounds of the approximation will be sharp as we increase the size of the domain. Heuristics for approximately counting satisfying assignments of propositional formulas exist, and toolboxes are readily available [10]. Finally, iterating through all the possible truth assignments to the formulas in the knowledge base can be avoided by sampling from the assignments using importance sampling (the idea is described in [11]).

We discussed the challenges of extending the framework to infinite domains, where the number of propositions is unbounded. Further examination of infinite domains is one of our future goals.

We only discussed modal logics with a single agent, since for multiple agents the number of non-equivalent epistemic situations even over a fixed set of propositions is unbounded. Another future goal is to explore if we can do inference efficiently in the multi agent setting despite the infinite state space.

## 8. ACKNOWLEDGMENTS

This research was supported by grants from ARO (W991NF-08-1-0242), ONR (N00014-11-10417), NSF (IIS-1012017), DOD (N00014-12-C-0263), and a gift from Intel.